\documentclass{iaus}
\usepackage{graphicx}

\title[The formation of H$_2$ and HD] %% give here short title %%
{The formation of H$_2$ and HD with the master equation approach}

\author[Biham, Lipshtat and Perets]   %% give here short author list %%
{Ofer Biham$^1$, 
Azi Lipshtat$^1$ 
\thanks{Present address: 
Department of Pharmacology and Biological Chemistry,
Mount Sinai School of Medicine,
New York, NY 10029, USA.
} 
\and Hagai B. Perets$^1$
\thanks{Present address: Faculty of Physics, 
Weizmann Institute of Science, Rehovot 76100, Israel.}
}
\affiliation{$^1$Racah Institute of Physics, The Hebrew University, 
Jerusalem 91904, Israel
\break email: biham@phys.huji.ac.il \\[\affilskip]
}

\pubyear{2005}
\volume{231}  %% insert here IAU Symposium No.
\pagerange{1-9}
\date{?? and in revised form ??}
\jname{Astrochemistry - Recent Successes and Current Challenges}
\editors{Dariusz C. Lis, Geoffrey A. Blake \& Eric Herbst, eds.}
\begin{document}

\maketitle

\begin{abstract}
The formation of H$_2$ and HD molecules on 
interstellar dust grains is studied using rate equation
and master equation models.
Rate equations are used in the analysis of 
laboratory experiments which examine the formation 
of molecular hydrogen on astrophysically
relevant surfaces.
However, under interstellar conditions, rate equations are
not suitable for the calculation of reaction rates
on dust-grain surfaces. 
Due to the low flux and the sub-micron
size of the grains, the populations of
H and D atoms on a single grain
are likely to be small. 
In this case the reaction rates are dominated by fluctuations 
and should be calculated using stochastic methods.
The rate of molecular hydrogen formation in interstellar
clouds is evaluated
using the master equation, taking into account
the distribution of grain sizes.

\keywords{ISM: molecules, ISM: abundances, molecular processes}
\end{abstract}

\firstsection % if your document starts with a section,

\section{Introduction}

The formation of molecular hydrogen in the interstellar medium 
is a process of fundamental importance
in astrophysics. 
H$_2$ molecules contribute to the cooling of clouds
during gravitational collapse and  
participate in
reactions that produce more complex molecules 
(\cite{Duley1984}; \cite{Williams1998}).
H$_2$ 
cannot form in the gas phase
efficiently enough to account for its observed abundance
(\cite{Gould1963}).  
It was thus proposed that
H$_2$ 
formation takes place on dust grains that act as
catalysts.
The process of H$_2$ formation on grains
can be broken up into several steps as follows.
An H atom approaching the surface of a grain has
a probability $\xi$ to stick and become adsorbed.
The adsorbed H atom (adatom)
resides on the surface for an average time 
$t_{\rm H}$ (residence time)
before it desorbs.
In the Langmuir-Hinshelwood mechanism,
the adsorbed H atoms quickly equilibrate 
and diffuse on the surface of the grain
either by thermal activation or by tunneling. 
When two adsorbed H atoms encounter each other, 
an H$_2$ molecule may form 
(\cite{Williams1968};\cite{Hollenbach1970};\cite{Hollenbach1971a};\cite{Hollenbach1971b};\cite{Smoluchowski1981};\cite{Aronowitz1985};\cite{Duley1986};\cite{Pirronello1988};\cite{Sandford1993};\cite{Takahashi1999};\cite{Farebrother2000}).
The steady state production rate of molecular hydrogen,
$R_{\rm H_2}$ (cm$^{-3}$ s$^{-1}$)
can be expressed by

\begin{equation}
    R_{{\rm H_2}} = {1 \over 2}
        n_{{\rm H}} v_{{\rm H}} \sigma \gamma n_{\rm g},
\label{eq:salpeter}
\end{equation}

\noindent
where
$n_{{\rm H}}$ (cm$^{-3}$)
and
$v_{{\rm H}}$ (cm s$^{-1}$)
are the number density and the speed
of H atoms in the gas phase, respectively,
$\sigma$ (cm$^{2}$) 
is the average cross-sectional area of a grain and
$n_{{\rm g}}$ (cm$^{-3}$)
is the number density of dust grains.
The parameter
$\gamma$ is the fraction of H atoms striking the grain
that eventually form a molecule, namely $\gamma = \xi \eta$,
where
$\eta$
is the probability that an H adatom on the surface
will recombine with another H atom
to form
H$_2$.
The probability $\xi$ for an H atom to become adsorbed on a grain surface
covered by an ice mantle was calculated by 
Buch and Zhang (1991) 
and 
Masuda et al. (1998).
They found that $\xi$ depends on the surface
temperature and on the energy of the irradiation beam.
For a surface at 10 K and beam temperature of 350 K,
Masuda et al. (1998) obtained a sticking coefficient around 0.5.

A simplified version of 
Eq.~(\ref{eq:salpeter})
is often used 
for the evaluation of the H$_2$ formation rate 
in models of interstellar chemistry.
It takes the form

\begin{equation}
R_{\rm H_2} = R \cdot n_{\rm H} n,
\label{eq:astronomers}
\end{equation}

\noindent
where
$R$ (cm$^3$ s$^{-1}$)
is the rate coefficient and
$n=n_{\rm H} + 2 n_{\rm H_2}$ 
(cm$^{-3}$) 
is the total density of hydrogen atoms
in both atomic and molecular ($n_{\rm H_2}$) form.
In previous studies it was assumed 
that the
total grain mass is 
about $1\%$ of the 
hydrogen mass in the cloud, with
a single grain size of
$r=0.17 \mu$m and 
a grain-mass density of 
$\rho_g = 2$ (gram cm$^{-3}$)
(\cite{Hollenbach1971b}).
Under these assumptions, for
gas temperature of 100 K
and $\gamma = 0.3$,
one obtains
$R \simeq 10^{-17}$ (cm$^3$ s$^{-1}$),
in agreement with observations for diffuse clouds
(\cite{Jura1975}).
More recent observations in photon dominated regions,
indicate that in these regions the rate coefficient
should be
$R \simeq 10^{-16}$ (cm$^3$ s$^{-1}$),
in order to account for the observed 
H$_2$ abundance
(\cite{Habart2003}).

\section{Laboratory Experiments}

In the last few years, the formation of molecular hydrogen 
on interstellar dust analogues 
was studied in a series of laboratory experiments
(\cite{Pirronello1997a};\cite{Pirronello1997b};\cite{Pirronello1999};\cite{Roser2002};\cite{Roser2003}).
In these experiments, beams of H and D atoms were irradiated
on the surface. 
The production of HD molecules
was measured both during irradiation and
during a subsequent temperature programmed desorption 
(TPD) experiment. 
In this experiment, the temperature of the sample is raised
quickly to either desorb particles (atoms and molecules) 
that got trapped on the surface or
to enhance their diffusion and 
subsequent reaction and desorption. 
The formation of molecular hydrogen
was studied on samples of polycrystalline olivine 
(\cite{Pirronello1997a};\cite{Pirronello1997b}),
amorphous carbon
(\cite{Pirronello1999})
and various types of amorphous ice samples,
namely high density ice (HDI) and low density ice (LDI) 
(\cite{Manico2001};\cite{Roser2002};\cite{Hornekaer2003}). 
The results of the TPD experiments, 
particularly on polycrystalline olivine and HDI, 
show second order kinetics for low coverage 
of H and D atoms on the surface.
In second order kinetics the peak shifts towards
higher temperature as the coverage is reduced.
This is due to the larger distancees that atoms 
need to diffuse before they encounter each other and recombine.
This indicates that on these samples
the surface mobility of H and D atoms 
is dominated by thermal hopping rather than tunneling.
If tunneling was dominant, molecules could form 
efficiently during irradiation, resulting in first
order kinetics.

The results of the TPD experiments were analyzed
using rate equation models 
(\cite{Katz1999};\cite{Cazaux2002};\cite{Cazaux2004};\cite{Perets2005a}).
The analysis involves fitting
the parameters of the rate equations
to the experimental TPD curves.
The most important parameters are
the activation energy barriers for 
diffusion and desorption of hydrogen atoms on the surface.
The models used by Katz et al. (1999) and Perets et al. (2005)
take into account only physisorption sites and thermal hopping
between them. 
The models of Cazaux and Tielens (2002;2004) also take into 
account tunneling and chemisorption sites.
However, the two analyses are in agreement on the fact that
tunneling between physisorption sites appears to be too slow to play an
important role in the TPD experiments
(\cite{Vidali2005a}).
Furthermore, recent experiments and calculations indicate that there are
energy barriers which are likely to prevent adsorbed H and D atoms
from entering chemisorption sites at these low temperatures.

Using the parameters obtained in the laboratory experiments,
the rate equations enabled the extrapolation from laboratory time
scales to astrophysical time scales.
Using rate equations, 
the efficiency of molecular hydrogen formation on these
surfaces under interstellar conditions 
and its dependence on the flux and grain temperature
were calculated.
It  was found that the
recombination efficiency strongly depends on the grain temperature.
Each sample exhibits 
a narrow window of high efficiency along the temperature axis.
For the polycrystalline olivine sample, this window is
between
$7 - 9$K,  
while for amorphous carbon it is between
$11 - 17$ K
(\cite{Vidali2005b}). 

\section{The Rate-Equation Model}
\label{sec:REModel}

Consider a diffuse interstellar cloud with
a density
$n_{\rm H}$
(cm$^{-3}$)
of H atoms,
and a density
$n_{\rm g}$
of dust grains.
The typical velocity 
$v_{\rm H}$ (cm s$^{-1}$)
of H atoms in the gas phase is
given by

\begin{equation}
v_{\rm H} = \sqrt{ {8 \over \pi} {k_B T_{\rm gas} \over m_{\rm H}} },
\label{eq:velocity}
\end{equation}

\noindent
where $m_{\rm H}=1.67 \cdot 10^{-24}$ (gram)
is the mass of an H atom
and $T_{\rm gas}$ is the gas temperature.
To evaluate the flux of atoms onto grain surfaces
we will assume, 
for simplicity,
that the grains are 
spherical with a
radius $r$ (cm).
The cross section of a grain is 
$\sigma = \pi r^2$
and its surface area is 
$4 \sigma$.
The density of adsorption sites on the surface
is $s$ (sites cm$^{-2}$)
and the number of adsorption sites on a grain is 
$S = 4 \pi r^2 s$.
The flux
$F_{\rm H}$ (atoms s$^{-1}$) 
of H atoms onto the surface of a single grain 
is given  by
$F_{\rm H} = n_{\rm H} v_{\rm H} \sigma$.

H atoms that collide and stick to the surface hop as random walkers
between adjacent sites until they either desorb 
or recombine into molecules.
The desorption rate of an H atom is 

\begin{equation}
W_{\rm H} =  \nu \cdot \exp (- E_{1} / k_{B} T),  
\label{eq:P1}
\end{equation}

\noindent
where $\nu$ is the attempt rate 
(typically assumed to be $10^{12}$ s$^{-1}$), 
$E_{1}$ 
is the activation energy barrier for desorption 
of an H atom and $T$ is the surface temperature.
The hopping rate of an H atoms is

\begin{equation} 
a_{\rm H} =  \nu \cdot \exp (- E_{0} / k_{B} T), 
\label{eq:Alpha}
\end{equation}

\noindent
where
$E_{0}$ is the activation energy barrier for H diffusion. 
Throughout this paper we use the parameters obtained
experimentally for amorphous carbon,
namely the activation energies are
$E_0=44.0$ meV
and
$E_1=56.7$ meV
(\cite{Katz1999}),
and the density of adsorption sites on the
surface is
$s \simeq 5 \times 10^{13}$ 
(sites cm$^{-2}$)
(\cite{Biham2001}).
For the
density of hydrogen atoms in the gas phase
we take
$n_{\rm H}=10$ 
(atoms cm$^{-3}$).
The gas temperature is taken as 
$T_{\rm gas}=90$ K,
thus 
$v_H=1.37 \times 10^5$ (cm s$^{-1}$). 
These are typical values for diffuse interstellar clouds.

The number of H atoms on the grain is 
denoted by
$N_{\rm H}$
and its expectation value
is denoted by
$\langle N_{\rm H} \rangle$.
The  rate 
$A_{\rm H} = a_{\rm H}/S$
is approximately the inverse of the time
$t_s$
required for an atom
to visit nearly all the
adsorption sites on the grain surface.
This is due to the fact that in two dimensions the 
number of distinct sites visited by a random walker
is linearly proportional to the number of steps, up
to a logarithmic correction
(\cite{Montroll1965}).

The formation of 
H$_2$ molecules on interstellar dust grains can be 
described by the rate equation

\begin{equation}
\label{eq:Ngrain}
{ {d{ \langle N_{\rm H} \rangle }} \over {dt}}  =  F_{\rm H} 
- W_{\rm H} \langle N_{\rm H} \rangle - 2 A_{\rm H} {\langle N_{\rm H} \rangle}^{2}. 
\label{eq:N1grain} \\
\end{equation}

\noindent
The first term on the right hand side describes the flux of H atoms,
the second term describes the desorption of H atoms and the third
term describes the diffusion and recombination. 
Here we assume, for simplicity, that all H$_2$ molecules desorb from the
surface upon formation.
The production rate 
$R_{\rm H_2}$ 
(cm$^{-3}$ s$^{-1}$)
of H$_2$ molecules is given by

\begin{equation}
R_{\rm H_2} = A_{\rm H} \langle N_{\rm H} \rangle^2 n_{\rm g}.
\label{eq:RH2}
\end{equation}

\noindent
Solving the rate equations under steady state conditions
for astrophysically relevant values of the flux and temperature,
it was found 
(\cite{Biham2002})
that molecular hydrogen formation
is efficient only 
within a narrow window of grain temperatures,
$T_0 < T < T_1$,
where

\begin{equation} 
T_0 = \frac{E_0}{k_{\rm B} \ln(\nu S/F_{\rm H})}
\end{equation}

\noindent
and

\begin{equation} 
T_1 = \frac{2 E_1 - E_0}{k_{\rm B} \ln(\nu S/F_{\rm H})}.
\end{equation}

\noindent
At temperatures below $T_0$ the
mobility of H atoms on the surface is very low,
sharply reducing the production rate.
At temperatures above $T_1$ most atoms quickly desorb
before they encounter each other and form molecules.

Rate equations are an ideal tool for the simulation of 
surface reactions, due to their simplicity and high computational efficiency.
In particular, they account correctly for the temperature dependence
of the reaction rates.
However, in the limit of small grains under low flux they become unsuitable.
This is because they ignore the fluctuations as
well as the discrete nature of the populations of atoms on the grain
(\cite{Charnley1997};\cite{Caselli1998};\cite{Shalabiea1998};\cite{Stantcheva2001}). 
For example, as the number of H atoms on a grain fluctuates 
in the range of 0, 1 or 2,
the H$_2$ formation rate cannot be obtained from the average 
number alone.
This can be easily understood, 
since the recombination process requires at least two
H atoms simultaneously on the surface.
In this limit, the master equation approach is needed in order
to evaluate the H$_2$ formation rate.

\section{The Master Equation}
\label{sec:MasterModel}

Consider a grain exposed to a flux of H atoms.
The probability that there are 
$N_{\rm H}$ hydrogen atoms  
on its surface 
is given by
$P_{\rm H}(N_{\rm H})$,
where
$N_{\rm H}=0, 1, 2,\dots,S$.
The master equation provides 
the time 
derivatives
of these probabilities.
It is a set of coupled linear differential equations
of the form

\begin{eqnarray}
\label{eq:Nmicro}
\dot P_{\rm H}(N_{\rm H}) &=&  
F_{\rm H} \left[ P_{\rm H}(N_{\rm H}-1) - P_{\rm H}(N_{\rm H}) \right] 
+ W_{\rm H} \left[ (N_{\rm H}+1) P_{\rm H}(N_{\rm H}+1) - N_{\rm H} P_{\rm H}(N_{\rm H}) \right] \nonumber \\
                &+& A_{\rm H} \left[ (N_{\rm H}+2)(N_{\rm H}+1) P_{\rm H}(N_{\rm H}+2) 
-  N_{\rm H}(N_{\rm H}-1) P_{\rm H}(N_{\rm H}) \right].   
\end{eqnarray}

\noindent
Each equation includes three terms.
The first term describes the effect of the incoming 
flux $F_{\rm H}$.
The probability $P_{\rm H}(N_{\rm H})$ increases 
when an H atom is adsorbed on a grain that already
has $N_{\rm H}-1$ adsorbed atoms 
[at a rate of $F_{\rm H} P_{\rm H}(N_{\rm H}-1)$], 
and decreases when a new atom is adsorbed on a grain with
$N_{\rm H}$ atoms on it
[at a rate of $F_{\rm H} P_{\rm H}(N_{\rm H})$].
The second term includes the effect of desorption. 
An H atom that is
desorbed from a grain with $N_{\rm H}$ atoms,  decreases the
probability $P_{\rm H}(N_{\rm H})$
[at a rate of
$N_{\rm H} W_{\rm H} P_{\rm H}(N_{\rm H})$], 
and increases the probability
$P_{\rm H}(N_{\rm H}-1)$
at the same rate.
The third term describes the effect of H$_2$ formation. 
The production of one molecule reduces the number of H atoms on the
surface from 
$N_{\rm H}$ to $N_{\rm H}-2$.
For one pair of H atoms the recombination rate is 
proportional to 
the sweeping rate
$A_{\rm H}$ 
multiplied by 2 since both atoms are mobile
simultaneously.
This rate is multiplied by
the number of possible pairs of atoms, namely
$N_{\rm H}(N_{\rm H}-1)/2$. 
Note that the equations for 
$\dot P_{\rm H}(0)$ 
and
$\dot P_{\rm H}(1)$ 
do not include all the terms, because at least one H 
atom is required for desorption to occur and at least two
for recombination.
The rate of formation of H$_2$ molecules on a single grain,
$R_{\rm H_2}^{\rm grain}$ (cm$^{-3}$ s$^{-1}$), 
is thus
given by 

\begin{equation}
R_{\rm H_2}^{\rm grain} = 
A_{\rm H} [\langle N_{\rm H}^2\rangle - \langle N_{\rm H}\rangle] 
\label{eq:Rgrain}
\end{equation}

\noindent
where

\begin{equation}
\langle N_{\rm H}^k \rangle = \sum_{N_{\rm H}=0}^{\infty} N_{\rm H}^k P_{\rm H}(N_{\rm H}).
\label{eq:defmomentk}
\end{equation}

\noindent
is the $k$th moment of the distribution 
$P_{\rm H}(N_{\rm H})$.

The time dependence of 
$R_{\rm H_2}^{\rm grain}$
can be obtained by numerically integrating Eqs.
(\ref{eq:Nmicro})
using a standard Runge-Kutta stepper.
For the case of steady state,
namely
$\dot P_{\rm H}(N_{\rm H})=0$
for all $N_{\rm H}$,
an analytical solution for 
$P_{\rm H}(N_{\rm H})$
was obtained,
in terms of
$A_{\rm H}/W_{\rm H}$
and
$W_{\rm H}/F_{\rm H}$
(\cite{Green2001};\cite{Biham2002}).
The rate of formation of H$_2$ molecules
on a single grain with $S$ adsorption sites
is given by
(\cite{Green2001};\cite{Biham2002})

\begin{equation}
R_{\rm H_2}^{\rm grain} = 
{F_{\rm H} \over {2}} 
{ {I_{W_{\rm H}/A_{\rm H}+1} \left(2\sqrt{2 F_{\rm H}/A_{\rm H}}\right)} 
\over 
{I_{W_{\rm H}/A_{\rm H}-1} \left(2\sqrt{2 F_{\rm H}/A_{\rm H}}\right)} },  
\label{eq:Rexact}
\end{equation}

\noindent
where $I_x(y)$ is the modified Bessel function.

In Fig. 1 we present the H$_2$ production rate on a single
grain of amorphous carbon vs. the grain size, which is
parametrized by the number of adsorption sites, $S$.  
The conditions, specified in Sec. 3, are typical in diffuse 
clouds and the grain temperature is $T= 18$ K. 
It is observed that for large grains the rate equation
results coincide with those of the master equation. 
However, for small grains the rate equation over-estimates the
production. The master equation accounts correctly for the
rate of H$_2$ formation on grains of all sizes.

\begin{figure}
\includegraphics[height=5.5in,width=4.5in,angle=270]{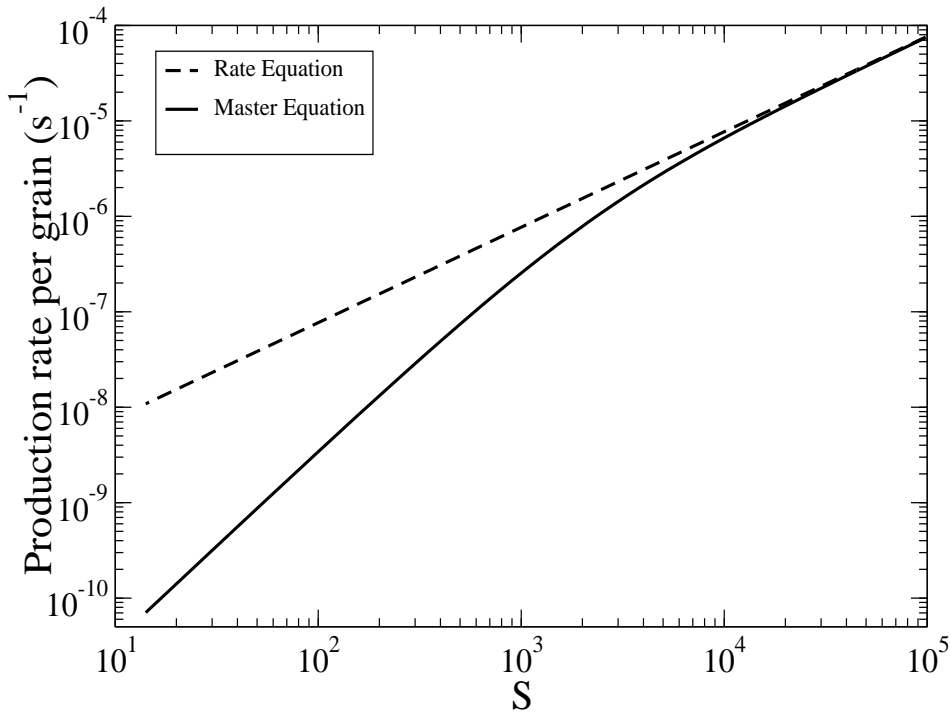}
\caption{The production rate of H$_2$ molecules on a single grain 
of amorphous carbon with $S$ adsorption sites.
For large grains the rate equation results (dashed line) are in agreement
with the master equation (solid line), while for small grains the
rate equation over-estimates the production.
The parameters are specified in Sec. 3.}
\label{fig:1}
\end{figure}

\section{Effects of the Grain Size Distribution}

Interstellar dust grains exhibit a broad size distribution
between several nano-meters and fractions of
a micron.
This distribution can be approximated by a power law of the form 
(\cite{Mathis1977};\cite{Weingartner2001}) 

\begin{equation}
n_g(r) = c r^{-\alpha} 
\label{eq:distribution} 
\end{equation}

\noindent
where 
$n_g(r) dr$ (cm$^{-3}$)
is the density of grains with radii in the range $(r,r+dr)$
in the cloud. 
This distribution is bounded between the upper cutoff
$r_{\rm max}= 0.25 \mu {\rm m}$ 
and the lower cutoff
$r_{\rm min} = 5 {\rm nm}$. 
The exponent is  
$\alpha \simeq 3.5$ 
(\cite{Draine1984}).
Clearly, this distribution is dominated by the small grains.
This fact has important implications on the formation rate
of molecular hydrogen. 
The total surface area is
much larger than in the case of a homogeneous population of grains
with radii of the order of
$0.25 \mu {\rm m}$,
with the same total mass density of about $1\%$
of the hydrogen mass density.
This tends to enhance the H$_2$ formation rate.
However, below some grain size the efficiency of H$_2$
formation is sharply reduced. 
Due to these competing effects the production of molecular
hydrogen should be integrated over the whole range
of grain sizes.

In Fig. 2 we present the rate coefficient
$R$ (cm$^3$ s$^{-1}$) vs. grain temperature
for the case of a power-law distribution
of grain sizes (solid line) and
for the case in which all grains are of radius
$r=0.17 \mu$m, with the same total mass of grains. 
It turns out that within the temperature window in which H$_2$
formation is efficient, 
the increase in the total surface area due to
the broad size distribution, enhances the
production rate by about an order of magnitude compared to
a homogeneous population of sub-micron size grains
(\cite{Lipshtat2005}).
This is in spite of the reduced efficiency of H$_2$ formation
on small grains.

\begin{figure}
\includegraphics[height=5.5in,width=4.5in,angle=270]{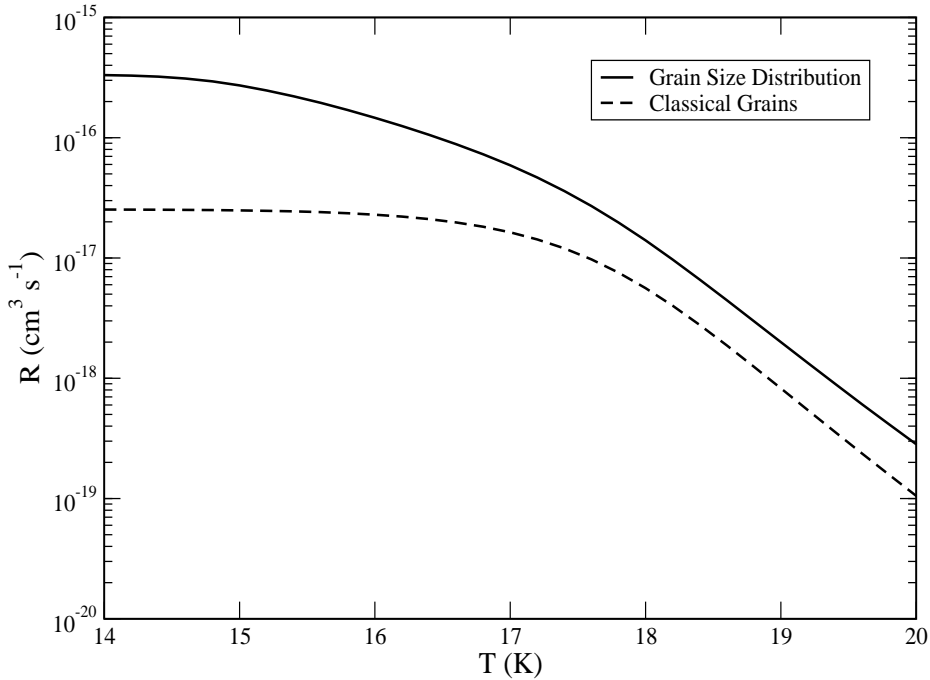}
\caption{The rate coefficient 
$R$ (cm$^3$ s$^{-1}$)
of H$_2$ formation vs. grain temperature, obtained from the
master equation for the case of a power law distribution
of grain sizes (solid line). The dashed line
shows the rate coefficient for the case in which the same
mass of grains consists of a homogeneous
population of ``classical grains'' of radius 
$r=0.17 \mu$m.} 
\label{fig:2}
\end{figure}

\section{Effects of Grain Porosity}

The laboratory experiments on interstellar dust analogues
have shown that H$_2$ formation on dust grains is efficient
within a range of grain temperatures below 20 K.
However, high abundances of H$_2$ have also been observed in warmer
clouds, including photon-dominated regions (PDRs), 
where grain temperatures may reach 50 K.
The possible effect 
of grain porosity on H$_2$ formation was recently
examined using a rate equation model
(\cite{Perets2005b}).
It is found that porosity extends the 
temperature range in which H$_2$ formation is efficient
towards higher temperatures. 
This is because H atoms that desorb from the internal surfaces of
the pores may re-adsorb many times and thus stay longer on the surface.
However, this extension is expected to be of only 
several degrees (K) and by itself
cannot account for the formation of molecular hydrogen
in PDRs.

Another effect of porosity that may be relevant
to molecular hydrogen formation is that
porous grains may be significantly
colder than predicted by models which use compact, 
spherical grains
(\cite{Blanco1980};\cite{Voshchinnikov1999}).  
While models of spherical grains predict grain temperatures in the 
range of $30 - 50$ K in PDRs,
the actual temperatures of porous interstellar grains
may thus be significantly lower. 
In the most favorable cases, the grain temperatures may go down to
the range of $15 - 30$ K for large grains and to $10 - 20$  
K for small grains 
(\cite{Blanco1980}). 

\section{HD Formation on Grains}

Unlike the H$_2$ molecules in the interstellar medium 
which do not form efficiently
in the gas phase, HD molecules have an efficient gas-phase
formation mechanism. 
It consists of a sequence of reactions in which
an ionized D atom reacts with an H$_2$ molecule to form HD.
The possibility that in diffuse clouds
some HD molecules are also formed on grain surfaces
was recently examined
(\cite{Lipshtat2004}).
Due to the lower abundance of D atoms vs. H atoms
(by a factor of about $10^{-5}$),
the master equation is required for such calculations
even on relatively large grains.
In case that D atoms bound more strongly to the surface 
than H atoms, the 
temperature window in which HD formation is efficient
is extended towards higher temperatures, compared to
the range of efficient H$_2$ formation.
The assumption that D atoms are more strongly bound to the grains
is plausible, but so far has not been confirmed experimentally.

\section{Complex Reaction Networks of Multiple Species}

The master equation can be generalized to describe complex reaction
networks of multiple species on grain surfaces.
Recently, it was applied to the study of the reaction
network that lead to the formation of
H$_2$CO and CH$_3$OH on grains
(\cite{Stantcheva2002};\cite{Stantcheva2003}).
Such surface reaction networks are dominated by H atoms
because hydrogen is the most abundant specie and also
because it reacts with most of the other atomic
and molecular species adsorbed on grain surfaces.
Furthermore, hydrogen atoms hop on the surface much faster than
other adsorbed species and thus quickly react with other species
even at very low coverages.
For such reaction networks, the dynamical variables in the master
equation are the probabilities
$P(N_{\rm H},N_{\rm O},N_{\rm OH},\dots)$
that there are $N_{\rm X}$ copies of specie X on the
grain.
Setting suitable cutoffs, each population of each specie is
limited to
$N_{\rm X}=0,1,\dots,N_{\rm X}^{\rm max}$. 
The problem is that as the number of reactive species
increases, the number of coupled equations in the master
equation increases exponentially, limiting the feasibility
of this approach. 
Using the fact that the reaction networks are usually sparse,
it is possible to dramatically reduce the number of equations
(\cite{Lipshtat2004}).
This makes the master equation applicable even for complex
surface reaction networks with a large number of reactive
species.

The master equation can be solved either by direct integration,
as described above, or by Monte Carlo (MC) methods
(\cite{Charnley2001}).
A significant advantage of direct integration is that the
equations can be easily coupled to the rate equations of
gas-phase chemistry. 
Furthermore, a single run of the direct integration method 
provides the entire 
probability distribution of populations of adsorbed species
for the simulated conditions as well as the reaction rates.
This is in contrast with the MC methods,
which typically require large computational resources
due to the need to accumulate much statistical information.

\section{Summary}

We have presented
the master equation approach for the evaluation of molecular
hydrogen formation rates on interstellar dust grains.
This approach is required in the limit of small grains and
low flux where the population of hydrogen atoms on
each grain is small. In this limit, the reaction rate is
dominated by fluctuations. Thus, the rate equation approach, which
incorporates the mean-field approximation fails.
Applications of the master equation approach for the
evaluation of the formation rates of H$_2$, HD as well
as more complex molecules are discussed. 

\begin{acknowledgments}
We thank E. Herbst, V. Pirronello and G. Vidali 
for helpful discussions.
This work was supported by the Israel Science Foundation
and the Adler Foundation for Space Research.
\end{acknowledgments}

\begin{discussion}

\discuss{Hornekaer}{Will the inclusion of chemisorption sites
on dust grains turn even small interstellar grains into efficient
contributors to H$_2$ formation in photon dominated regions (PDRs) due to
the fact that thermal spikes could activate recombination of chemisorbed
H atoms?}

\discuss{Biham}{In our analysis so far we have not included fluctuations
in grain temperatures. Such fluctuations are expected to be significant 
only for grains of radii smaller than about 10 nm.
Chemisorbed H atoms are expected to contribute to H$_2$ formation
when grain temperatures reach several hundred degrees (K).
However, a careful analysis is required in order to quantify the
contribution of such mechanism.
}

\discuss{Aikawa}{You concluded that H migrates by thrmal hopping,
while Cazaux fits the same experimental data assuming tunneling. Would
you tell me why you need to exclude tunneling and why you and Cazaux
reached different conclusions on H-migration?}

\discuss{Biham}{The temperature programmed desorption (TPD) curves obtained
for the polycrystalline olivine sample (Katz et al. (1999) exhibit
second order kinetics at lower coverage, namely, as the
coverage decreases, the TPD peak shifts to higher temperatures.
This indicates that the
HD molecules are formed by thermal hopping during the TPD run.
If tunneling alone provided sufficient mobility
for H atoms to diffuse and recombine during irradiation, when the 
surface temperature is around 6 K, all TPD curves would be of first order
kinetics. For other samples, such as the amorphous carbon, the evidence
for second order kinetics is not as strong and tunneling is not ruled
out completely.}

\end{discussion}


\begin{thebibliography}{10}

\bibitem[Aronowitz \& Chang 1985]{Aronowitz1985}
Aronowitz S., \& Chang S.
1985, ApJ, 293, 243

\bibitem[Biham et al. 2001]{Biham2001}
Biham, O., Furman, I., Pirronello, V., Vidali, G.
2001 ApJ, 553, 595.

\bibitem[Biham \& Lipshtat 2002]{Biham2002}
Biham, O. \& Lipshtat A. 
2002, Phys. Rev. E 66, 055103 

\bibitem[Blanco \& Bussoletti 1980]{Blanco1980}
Blanco A., Bussoletti B.
1980, Astrophys. Space Sci., 67, 105

\bibitem[Buch \& Zhang 1991]{Buch1991}
Buch, V. \& Zhang, Q.
1991, ApJ, 379, 647

\bibitem[Cazaux \& Tielens 2002]{Cazaux2002}
Cazaux, S. \& Tielens, A. G. G. M.
2002, ApJ, 575, L29

\bibitem[Cazaux \& Tielens 2004]{Cazaux2004}
Cazaux, S. \& Tielens, A. G. G. M.
2004, ApJ, 604, 222

\bibitem[Charnley et al. 1997]{Charnley1997}
Charnley, S.B., Tielens A.G.G.M., Rodgers, S.D.
1997, ApJ, 482, L203

\bibitem[Charnley 2001]{Charnley2001}
Charnley, S.B.
2001, ApJ, 562, L99

\bibitem[Caselli et al. 1998]{Caselli1998}
Caselli, P., Hasegawa, T.I. \& Herbst, E.
1998, ApJ, 495, 309

\bibitem[Draine \& Lee 1984]{Draine1984}
Draine B.T. \&  Lee, H.M. 
1984, ApJ, 285, 89

\bibitem[Duley \& Williams 1984]{Duley1984}
Duley, W.W. \& Williams, D.A.
1984, Interstellar Chemistry (Academic Press, London)

\bibitem[Duley \& Williams 1986]{Duley1986}
Duley, W.W. \& Williams, D.A.
1986, MNRAS, 223, 177

\bibitem[Farebrother et al. 2000]{Farebrother2000}
Farebrother, A.J., Meijer A.J.H.M., Clary D.C., Fisher A.J.
2000, Chem. Phys. Lett., 319, 303

\bibitem[Gould \& Salpeter 1963]{Gould1963}
Gould, R.J. \& Salpeter, E.E.
1963, ApJ., 138, 393

\bibitem[Green et al. 2001]{Green2001}
Green N.J.B., Toniazzo T., Pilling M.J. Ruffle, D.P. Bell, N., 
Hartquist T.W. 
2001, A\&A, 375, 1111 

\bibitem[Habart et al. 2003]{Habart2003}
Habart, E., Boulanger, F., Verstraete, L., Pineau des For\^ets, G.,  
Falgarone, E., Abergel, A.
2003, A\&A, 397, 623

\bibitem[Hollenbach \& Salpeter 1970]{Hollenbach1970}
Hollenbach, D. \& Salpeter, E.E.
1970, J. Chem. Phys., 53, 79

\bibitem[Hollenbach \& Salpeter 1971]{Hollenbach1971a}
Hollenbach, D. \& Salpeter, E.E.
1971, ApJ, 163, 155

\bibitem[Hollenbach et al. 1971]{Hollenbach1971b}
Hollenbach, D., Werner M.W., Salpeter, E.E.
1971, ApJ, 163, 165

\bibitem[Hornekaer et al.2003]{Hornekaer2003}
Hornekaer, L., Baurichter, A., Petrunin, V. V., Field, D., Luntz, A. C.
2003, Science, 302, 1943

\bibitem[Jura 1975]{Jura1975}
Jura, M. 
1975, ApJ, 197, 575

\bibitem[Katz et al. 1999]{Katz1999}
Katz, N., Furman, I., Biham, O., Pirronello V., Vidali, G.
1999, ApJ, 522, 305

\bibitem[Lipshtat et al. 2004]{Lipshtat2004}
Lipshtat, A., Biham, O., Herbst E.,
2004, MNRAS, 348, 1055

\bibitem[Lipshtat \& Biham 2005]{Lipshtat2005}
Lipshtat, A. \& Biham, O.
2005, MNRAS, 362, 666

\bibitem[Manico et al. 2001]{Manico2001}
Manico, G., Raguni, G., Pirronello, V., Roser, J.E., Vidali G.
2001, ApJ, 548, L253

\bibitem[Masuda et al. 1998]{Masuda1998}
Masuda, K., Takahashi, J., Mukai, T. 
1998, A\&A, 330, 773

\bibitem[Mathis et al. 1977]{Mathis1977}
Mathis J.S., Rumpl W., Nordsieck K.H. 
1977, ApJ, 217, 425

\bibitem[Montroll \& Weiss 1965]{Montroll1965}
Montroll, E.W. \& Weiss, G.H. 
1965, J. Math. Phys., 6, 167

\bibitem[Perets et al. 2005]{Perets2005a}
Perets, H.B., Biham, O., Manic\'o, G., Pirronello, V.,
Roser, J., Swords, S., Vidali, G.
2005, ApJ, 627, 850

\bibitem[Perets \& Biham 2005]{Perets2005b}
Perets, H.B. \& Biham, O.
2005, preprint

\bibitem[Pirronello \& Averna 1988]{Pirronello1988}
Pirronello, V. \& Averna D. 
1988, A\&A, 201, 196

\bibitem[Pirronello et al. 1997a]{Pirronello1997a}
Pirronello, V., Liu, C., Shen L., Vidali, G.
1997a, ApJ, 475, L69

\bibitem[Pirronello et al. 1997b]{Pirronello1997b}
Pirronello, V., Biham, O., Liu, C., Shen L., Vidali, G.
1997b, ApJ, 483, L131

\bibitem[Pirronello et al. 1999]{Pirronello1999}
Pirronello, V., Liu, C., Roser J.E, Vidali, G.
1999, A\&A, 344, 681

\bibitem[Roser et al. 2002]{Roser2002}
Roser, J.E., Manico, G., Pirronello, V., Vidali, G.
2002, ApJ., 581, 276

\bibitem[Roser et al. 2003]{Roser2003}
Roser, J.E., Swords S., Vidali, G.
2003, ApJ. 596, L55

\bibitem[Sandford \& Allamandolla 1993]{Sandford1993}
Sandford S.A., \& Allamandolla L.J.
1993, ApJ, 409, L65

\bibitem[Shalabiea et al. 1998]{Shalabiea1998}
Shalabiea O.M., Caselli P., Herbst E. 
1998, ApJ, 502, 652

\bibitem[Smoluchowski 1981]{Smoluchowski1981}
Smoluchowski, R.
1981, Astrophys. Space Sci., 75, 353

\bibitem[Stantcheva et al. 2001]{Stantcheva2001}
Stantcheva T., Caselli P., Herbst E. 
2001, A\&A, 375, 673

\bibitem[Stantcheva et al. 2002]{Stantcheva2002}
Stantcheva T., Shematovich V.I.,  Herbst E., 2002, 
A\&A, 291, 1069

\bibitem[Stantcheva \& Herbst 2003]{Stantcheva2003}
Stantcheva T. \& Herbst E. 
2003, MNRAS, 340, 983

\bibitem[Takahashi et al. 1999]{Takahashi1999}
Takahashi, J, Masuda, K., Nagaoka, M. 
1999, MNRAS, 306, 22

\bibitem[Vidali et al. 2005]{Vidali2005a}
Vidali, G., Roser, J., Manic\'o, G., Pirronello, V.,
Perets, H.B. and Biham, O., 
Proceedings of the International Symposium on Light, Dust and
Chemical Evolution, 
Gerace, Italy
(Sept. 26-30, 2004),
2005, Journal of Physics: Conference Series, 6, 36 

\bibitem[Vidali et al. 2005]{Vidali2005b}
Vidali, G., Roser, J., Manic\'o, G., Pirronello, V.
2005, IAU Sympusium 231, this proceedings

\bibitem[Voshchinnikov et al. 1999]{Voshchinnikov1999}
Voshchinnikov N.V., Semenov D.A., Henning T. 
1999, A\&A, 349, L25

\bibitem[Weingartner \& Draine 2001]{Weingartner2001}
Weingartner J.C. \& Draine B.T. 
2001, ApJ, 548, 296 

\bibitem[Williams 1968]{Williams1968}
Williams, D.A.
1968, ApJ, 151, 935

\bibitem[Williams 1998]{Williams1998}
Williams, D.A.
1998, Faraday Discussions, 109, 1

\end{thebibliography}
\end{document}